\documentclass{article}[11pt]
\usepackage{a4wide}
\usepackage{bm}
\usepackage{bbm}
\usepackage{epsfig,graphics}
\usepackage{amsmath,amssymb,amsfonts}
\usepackage{color}
\usepackage{epstopdf}
\usepackage[
colorlinks=true,
linkcolor=black,
breaklinks=true,
urlcolor=blue,
citecolor=green]{hyperref}

\newcommand{\be}{\begin{equation}}
\newcommand{\ee}{\end{equation}}
\newcommand{\ba}{\begin{eqnarray}}
\newcommand{\ea}{\end{eqnarray}}

\newcommand{\al}{&\!\!\!}

\newcommand{\order}[1]{\mathcal{O}\left(#1\right)}

\newcommand{\Tr}[1]{\left\langle #1 \right\rangle}
\newcommand{\mM}{\mathcal{M}}

\begin{document}

\title{ Cumulants of the QCD topological charge distribution}

\author{Feng-Kun Guo$^{1,}$\footnote{Email address:
      \texttt{fkguo@hiskp.uni-bonn.de} }~ and
      Ulf-G. Mei\ss ner$^{1,2,}$\footnote{Email address:
      \texttt{meissner@hiskp.uni-bonn.de} } \\[2mm]
      {\it\small$^1$Helmholtz-Institut f\"ur Strahlen- und Kernphysik and Bethe
      Center for Theoretical Physics,}\\
      {\it\small Universit\"at Bonn, D-53115 Bonn, Germany}\\
      {\it\small$^2$Institute for Advanced Simulation, Institut f\"{u}r
       Kernphysik and J\"ulich Center for Hadron Physics,}\\
      {\it\small Forschungszentrum J\"{u}lich, D-52425 J\"{u}lich, Germany}
}
\date{\today}

\maketitle

\begin{abstract}

  The distribution of the QCD topological charge can be described by
  cumulants, with the lowest one being the topological susceptibility.
  The vacuum energy density in a $\theta$-vacuum is the generating function for
  these cumulants. In this paper, we derive the vacuum energy density in SU(2)
  chiral perturbation theory up to next-to-leading order keeping different up
  and down quark masses, which can be used to calculate any cumulant of the
  topological charge distribution. We also give the expression for the case
  of SU($N$) with degenerate quark masses. In this case, all cumulants depend on
  the same linear combination of low-energy constants and chiral logarithm, and
  thus there are sum rules between the $N$-flavor quark condensate
  and the cumulants free of next-to-leading order corrections.

\end{abstract}

\medskip

\newpage

\section{Introduction}

Because of the axial U(1) anomaly, there exists a $\theta$-term in
quantum chromodynamics (QCD) which is a topological term. The partition function
of QCD in a $\theta$-vacuum is given by
\begin{equation}
  Z(\theta) = \int [D G][Dq][D\bar q]\, e^{-S_\text{QCD}[G,q,\bar q]-i\theta
  Q},
\end{equation}
where $S_\text{QCD}[G,q,\bar q]$ is the QCD action  at $\theta=0$ with $G$ and
$q$ being the gluon and quark fields, respectively, and $Q$ is the
topological charge
\begin{equation}
  Q = \frac1{32\pi^2} \epsilon_{\mu\nu\rho\sigma} \int d^4x\, G^{\mu\nu}(x)
  G^{\rho\sigma}(x)~,
\end{equation}
with $G^{\mu\nu}(x)$ the gluon field strength tensor.
In the Euclidean space with
a finite space-time volume $V$, the partition function $Z(\theta)$ is dominated
by the ground state, i.e. vacuum, energy of QCD for large enough $V$ (see,
e.g. Ref.~\cite{Brower:2003yx}), and we have
\begin{equation}
  Z(\theta) = e^{-V e_\text{vac}(\theta)}, \qquad\text{or}\qquad
  e_\text{vac}(\theta) = -\frac1{V} \ln Z(\theta)\, ,
\end{equation}
where $e_\text{vac}(\theta)$ is the vacuum energy density in the
$\theta$-vacuum.
The distribution of the topological charge can be described in terms of moments,
which are the expectation values $\Tr{Q^{2n}}_{\theta=0}$ with positive
integer $n$, or cumulants defined as
\begin{equation}
  c_{2n} = \left. \frac{d^{2n} e_\text{vac}(\theta)}{d\,\theta^{2n}}
  \right|_{\theta=0} \, .
  \label{eq:c2n1}
\end{equation}
The leading cumulant is the topological susceptibility, $c_2=\chi_t$. It and the
fourth cumulant are given by the well-known formulae
\begin{equation}
  \chi_t = \frac1{V} {\Tr{Q^2}_{\theta=0} }\,\qquad
  c_4 = -\frac1{V} \left( \Tr{Q^4} -3 \Tr{Q^2}^2 \right)_{\theta=0} \, .
\end{equation}
These topological quantities are important to understand the QCD vacuum as well
as to extract physical observables from lattice simulations at a fixed
topology~\cite{Brower:2003yx,Aoki:2007ka}.
They can be measured on lattice using various methods, see, e.g.,
Refs.~\cite{Gockeler:1986ge,Campostrini:1988cy,DelDebbio:2002xa,
D'Elia:2003gr,Durr:2006ky, Aoki:2007pw,
Chiu:2008kt,Horsley:2008gv,Giusti:2007tu,Giusti:2008vb,
Luscher:2010ik, Bazavov:2010xr,Bonati:2013tt, 
Cichy:2013rra,Bruno:2014ova,Cichy:2014qta,Bonati:2015uga}.

For large volume and small quark masses, the strong interaction dynamics is
determined by the Goldstone bosons originating from the spontaneous breaking of 
the
light-quark chiral symmetry, and thus can be well described by chiral
perturbation theory (CHPT)~\cite{Gasser:1983yg,Gasser:1984gg}. Both of $\chi_t$
and $c_4$ have been calculated in CHPT in both leading order (LO) and
next-to-leading order
(NLO)~\cite{Leutwyler:1992yt,Lenaghan:2001ur,Brower:2003yx,Aoki:2009mx,
Mao:2009sy,Bernardoni:2010nf,Bernard:2012fw,Bernard:2012ci}. Earlier discussions
in the large $N_c$ limit can be found in
Refs.~\cite{Witten:1979vv,Veneziano:1979ec}. The NLO calculations for $\chi_t$
in Refs.~\cite{Mao:2009sy,Bernard:2012ci} and for $c_4$ in
Ref.~\cite{Bernard:2012ci} were performed for an arbitrary number of flavors
with different masses, and based on the generating functionals of
CHPT~\cite{Gasser:1984gg} expanded around $\theta=0$ up to 2-point loops (up to
1-point tadpole loops for the topological susceptibility~\cite{Mao:2009sy}).

In this paper, we will derive a general formula for the vacuum energy density in
SU(2) chiral perturbation theory keeping different masses for the up and down
quarks. The derivation involves a direct calculation of the logarithm of the
determinant for the free Goldstone bosons in a $\theta$-vacuum, and thus does
not require an expansion up to a finite $n$-point loops. In this sense, it
contains a summation of all one-loop diagrams at NLO in the chiral expansion,
i.e. $\order{p^4}$ with $p$ denoting a small momentum or Goldstone boson mass,
contributing to the vacuum energy.
The expression for the vacuum energy density can then be used to calculate
{\em any} cumulant of the distribution of the QCD topological charge defined in
Eq.~\eqref{eq:c2n1}.

It was emphasized in Ref.~\cite{Bernard:2012ci} that lattice simulations of these
topological quantities with degenerate quarks are very interesting to pin down
the $N$-flavor quark condensate. Although both the topological susceptibility
and the fourth cumulant depend on several low-energy constants (LECs) in the NLO
chiral Lagrangian, in addition to the quark condensate, the authors found
an interesting linear combination, $\chi_t + N^2 c_4/4$ with $N$ the number of
flavors, independent of any LEC. Thus, such a combination is particularly
suitable for extracting the $N$-flavor averaged quark condensate whose absolute
value is
\begin{equation}
\Sigma_N = F_N^2 B_N\, ,
\end{equation}
where $F_N$, the pion decay constant, and $B_N$ are defined in the chiral limit.
For determinations of the quark
condensate from lattice calculations of the topological susceptibility, we refer
to Ref.~\cite{Bernardoni:2010nf,Cichy:2013rra,Bruno:2014ova}.
Stimulated by this insight, we will also
derive general expressions for the SU($N$) vacuum energy density and cumulants
with degenerate quarks. It turns out that all the cumulants depend on the same
linear combination of the NLO LECs and chiral logarithm. As a result, one can
construct linear combinations of the cumulants free of NLO corrections.

At this point, we notice that higher cumulants can be obtained from lower ones
and moments using the following recursion relation
\begin{equation}
   c_{2n} = (-1)^{n+1} \left[ \frac{\Tr{Q^{2n}} }{V} + \sum_{m=1}^{n-1} (-1)^{m}
  \binom{2n-1}{2m-1} \Tr{Q^{2(n-m)}}  c_{2m} \right]_{\theta=0} .
\end{equation}

\section{Vacuum energy in SU(2) chiral perturbation theory}

\subsection{Leading order}

Because the $\theta$-angle can be rotated to the phase of the quark mass matrix
by an axial U(1) rotation, the $\theta$-dependence of physical quantities can be
studied by using a complex quark mass matrix. At LO, $\order{p^2}$, of SU($N$)
chiral perturbation theory, the vacuum energy density in a $\theta$-vacuum for $N$
quarks is given by
\begin{equation}
e_\text{vac}^{(2)}(\theta) = - \frac{F_N^2}{4} \Tr{ \chi_\theta\, U_0^\dag +
\chi_\theta^\dag\, U_0},
\end{equation}
where $\chi_\theta = 2 B_N
\mM\,\exp(i\theta/N)$ with $\mM$ being the real and diagonal quark mass
matrix, and the vacuum alignment $U_0$ can be parametrized as a diagonal
matrix $U_0 = \text{diag}\{ e^{i\varphi_1}, e^{i\varphi_2}, \ldots,
e^{i\varphi_N} \} $ with the constraint $\sum_i \varphi_i = 0$. The angles
$\varphi_i$ are determined by minimizing the vacuum energy. It is equivalent
to removing the tree-level tadpole terms of the neutral Goldstone bosons which
would induce vacuum
instability~\cite{Dashen:1970et,Crewther:1979pi,Mereghetti:2010tp}.

In this section, we will study the case with $N=2$. We will drop the subscripts
in $F_2$ and $B_2$ to be consistent with the traditional notation in CHPT. With
$U_0 = \text{diag}\{ e^{i\varphi}, e^{-i\varphi}\} $, we have
\begin{equation}
  e_\text{vac}^{(2)}(\theta) = 2F^2 B\bar m \left( \cos\frac\theta2 \cos\varphi
  - \epsilon \sin\frac\theta2 \sin\varphi\right) ,
  \label{eq:eSU2LO}
\end{equation}
where $\bar m = (m_u+m_d)/2$ is the average mass of the up and down quarks and
$\epsilon=(m_d-m_u)/(m_u+m_d)$ quantifies the strong isospin breaking. Minimizing the
vacuum energy with respect to $\varphi$, one gets~\cite{Brower:2003yx}
\begin{equation}
  \tan \varphi = - \epsilon \tan\frac{\theta}{2}\, .
  \label{eq:phi}
\end{equation}
Substituting this into Eq.~\eqref{eq:eSU2LO}, we get the vacuum energy density
at LO, up to an additive normalization constant~\cite{Brower:2003yx}
\begin{equation}
  e_\text{vac}^{(2)}(\theta) = - F^2 \mathring M^2(\theta) \, ,
\end{equation}
where $\mathring{M}^2(\theta)$ is the LO pion mass squared
in a $\theta$-vacuum~\cite{Brower:2003yx}
\begin{equation}
 \mathring{M}^2(\theta) = 2 B \bar m \cos
\frac{\theta}{2} \, \sqrt{1 + \epsilon^2
 \tan^2\frac{\theta}{2}} \, .
 \label{eq:mpiLO}
\end{equation}
Notice that in the absence of the electromagnetic interaction, the neutral and
charged pions have the same mass at LO. The cumulants of the distribution of the
topological charge can then be easily obtained. For instance, the topological
susceptibility and the fourth cumulant at LO are
\begin{eqnarray}
 \chi_t^{(2)} = \frac12 F^2 B \bar m \left(1-\epsilon^2\right),\qquad
 c_4^{(2)} = -\frac18 F^2 B \bar m \left( 1 + 2\epsilon^2 - 3\epsilon^4 \right)
,
\end{eqnarray}
which have been derived before in
Refs.~\cite{Leutwyler:1992yt,Mao:2009sy,Aoki:2009mx}.

\subsection{Next-to-leading order}
\label{sec:su2NLO}

At NLO, there are contributions from both the tree-level terms in the
$\order{p^4}$ chiral Lagrangian and one-loop diagrams.
The vacuum energy density up to NLO is
given by
\begin{eqnarray}
  e_\text{vac}(\theta) = e_\text{vac}^{(2)}(\theta) +
  e_\text{vac}^{(4,\text{loop})}(\theta) +
  e_\text{vac}^{(4,\text{tree})}(\theta) \, ,
\end{eqnarray}
where $e_\text{vac}^{(2)}(\theta)$ is given in Eq.~\eqref{eq:eSU2LO},
$e_\text{vac}^{(4,\text{loop})}(\theta)$ is the one-loop contribution to be
calculated later on, and the NLO tree-level contribution is
\begin{eqnarray}
  e_\text{vac}^{(4,\text{tree})}(\theta) \al=\al
  - \frac{l_3}{16} \Tr{ \chi_\theta^\dag\, U_0 +
  \chi_\theta\, U_0^\dag }^2 + \frac{l_7}{16} \Tr{\chi_\theta^\dag\, U_0 -
  \chi_\theta\, U_0^\dag }^2
  \nonumber\\ \al\al
  - \frac{h_1+h_3}{4} \Tr{\chi_\theta^\dag\, \chi_\theta} -
  \frac{h_1-h_3}{2} \text{Re}\,(\text{det}\,\chi_\theta )
   \nonumber\\
   \al=\al  - \mathring{M}^4(\theta)   \left\{ l_3 + l_7 \left[\frac{(1-
   \epsilon^2) \tan(\theta/2)}{ 1 + \epsilon^2\tan^2(\theta/2) } \right]^2
   \right\}
  \nonumber\\ \al\al
    - 2  B^2 \bar m^2 \left[(h_1+h_3) \left(1+\epsilon^2\right) + (h_1
   - h_3) \left(1 - \epsilon^2\right) \cos\theta \right] ,
  \label{eq:e4tree}
\end{eqnarray}
where $l_3,l_7$ and $h_1,h_3$ are the LECs and high-energy constants (HECs),
respectively, in the NLO two-flavor chiral
Lagrangian~\cite{Gasser:1983yg},\footnote{Here we use the SU(2)$\times$SU(2) notation rather than the O(4) one in the original
paper, see, e.g., \cite{Knecht:1997jw}.}
and we have used Eq.~\eqref{eq:phi}.\footnote{In principle, the vacuum
alignment determined by minimizing the LO vacuum energy gets shifted due to the
presence of the higher order terms, $l_7$ in this case. However, this shift only
provides a perturbation and is of one order higher compared to the angle
$\varphi$ in Eq.~\eqref{eq:phi}. It introduces CP-odd vertices (see, e.g.,
Refs.~\cite{Bsaisou:2012rg,Bsaisou:2014oka}) and does not affect CP-even
quantities up to $\order{p^4}$, thus irrelevant for us.
It is for this reason that the topological susceptibility up to NLO in the
chiral expansion calculated in Ref.~\cite{Bernard:2012ci} agrees with that in
Ref.~\cite{Mao:2009sy}, where the vacuum alignment was calculated by minimizing
the LO and NLO vacuum energy, respectively.} Because both $l_3$ and $h_1$ are
ultraviolet (UV) divergent~\cite{Gasser:1983yg},
\begin{equation}
  l_3 = l_3^r -\frac{\lambda}{2}\, , \qquad h_1 = h_1^r + 2 \lambda \, ,
\end{equation}
with $\lambda$ the divergence at the space-time dimension $d=4$ in dimensional
regularization,
\begin{equation}
  \lambda = \frac{\mu^{d-4}}{16\pi^2} \left\{ \frac1{d-4} - \frac12 \left[
  \ln(4\pi) + \Gamma'(1) +1 \right] \right\} ,
\end{equation}
where $\mu$ is the scale in dimensional regularization,
$e_\text{vac}^{(4,\text{tree})}(\theta)$ is UV divergent as well, and the
divergence is (the divergence is nontrivial in a $\theta$-vacuum noticing the
$\theta$-dependence)
\begin{equation}
  e_\text{vac}^{(4,\text{tree},\infty)}(\theta) = - \frac{3\lambda}{2}
  \mathring{M}^4(\theta) \, .
  \label{eq:etreeinf}
\end{equation}
As will be shown, this divergence is exactly cancelled by the one from loops in
$e_\text{vac}^{(4,\text{loop})}(\theta)$.

Before proceeding to calculating the loop contribution to the vacuum energy
density, let us discuss the main difference
between our treatment (see below) and the one in
Refs.~\cite{Mao:2009sy,Bernard:2012ci}. In those papers the authors took
the expression of the generational functional in Ref.~\cite{Gasser:1984gg}. It is normalized to the free fields at
$\theta=0$ (notice that Refs.~\cite{Gasser:1983yg,Gasser:1984gg} assume
$\theta=0$). Then the loops were calculated using the Goldstone boson masses at
$\theta=0$, and the $\theta$-dependence is kept in the operator $\sigma^\chi$
defined as (we have replaced $U$ containing quantum fluctuations of Goldstone
bosons by $U_0$ relevant for the vacuum energy)
\begin{equation}
  \sigma^\chi_{PQ} = \frac18 \Tr{ \left\{ \lambda_P, \lambda_Q^\dag \right\}
  \left( \chi_\theta^\dag\, U_0 +  \chi_\theta\, U_0^\dag\right) } - \delta_{PQ}
  \mathring{M}_P^2(0) \, ,
  \label{eq:sigma}
\end{equation}
where $\lambda_P$ are linear combinations of the SU($N$) generators introduced
to diagonalize the LO mass term~\cite{Gasser:1984gg}, and $\mathring{M}_P(0)$
are the LO Goldstone boson masses at $\theta=0$.
This amounts to an expansion around $\theta=0$, which
is perfectly fine for the calculation of the cumulants of the topological
charge. However, we notice that the first term in the above equation is in fact
$\delta_{PQ} \mathring M^2_P(\theta)$. If we expand the one-loop generating
functional around the one for the free fields in a $\theta$-vacuum,
\begin{equation}
  Z_0(\theta) = \frac{i}{2}\, \ln \text{det}\, D_0(\theta) = \frac{i}{2}\,
  \text{Tr} \ln D_0(\theta) \, ,
\end{equation}
where $\text{Tr}$ stands for taking trace in both the flavor (this is the space
of the adjoint representation which is 3-dimensional for the SU(2) case) and
coordinate spaces, and $D_0(\theta)$ is a differential operator,
\begin{equation}
  D_{0\, PQ}(\theta) = \delta_{PQ} \left[ \partial_\mu \partial^\mu +
  \mathring{M}^2_P(\theta) \right] ,
  \label{eq:D0}
\end{equation}
 then $\mathring
M^2_P(0)$ in Eq.~\eqref{eq:sigma} needs to be replaced by $\mathring
M^2_P(\theta)$ and $\sigma^\chi_{PQ}$ vanishes. As a result, the only term left
in the one-loop generating functional relevant for the vacuum energy is
$Z_0(\theta)$. Thus, the vacuum energy density is given by
\begin{equation}
  e_\text{vac}^{(4,\text{loop})}(\theta) =  - \frac{i}{2\, V}\, \text{Tr} \ln
  D_0(\theta) \, .
\end{equation}

For the case of SU(2), because the neutral and charged pions have the same mass
at LO, $\mathring M^2_P(\theta)$ is given by Eq.~\eqref{eq:mpiLO}, and
$D_0(\theta) = \mathbbm{1}_{3\times3}\, [ \partial_\mu \partial^\mu +
  \mathring{M}^2(\theta) ]$, where the unit matrix has the dimension
of the adjoint representation for SU(2). Extending these considerations to the 
case of $N$
  degenerate quark flavors and using dimensional regularization, we obtain
\begin{eqnarray}
  e_\text{vac}^{(4,\text{loop})}(\theta) \al=\al -\frac{i}{2} (N^2-1) \int
  \frac{d^dp}{(2\pi)^d}\, \ln \left[-p^2+\mathring{M}^2(\theta)\right]
  \nonumber\\
  \al=\al \frac{i}{2} (N^2-1) \int
  \frac{d^dp}{(2\pi)^d}\int_0^\infty \frac{d\tau}{\tau} \,
  e^{-\tau\left[-p^2+\mathring{M}^2(\theta)\right]} \nonumber\\
  \al=\al (N^2-1)\, \mathring{M}^4(\theta) \left\{
   \frac{\lambda}{2} - \frac1{128\pi^2} \left[ 1 -2 \ln
   \frac{\mathring{M}^2(\theta)}{\mu^2} \right] \right\} .
   \label{eq:eloop}
\end{eqnarray}
For $N=2$, one sees that the UV divergence cancels exactly the one in
Eq.~\eqref{eq:etreeinf}. The sum of Eqs.~\eqref{eq:eSU2LO}, \eqref{eq:e4tree}
and \eqref{eq:eloop} provides the vacuum energy density in a $\theta$-vacuum up
to NLO,
\begin{equation}
  e_\text{vac}(\theta)
   = -F^2 \mathring M^2(\theta) - \mathring{M}^4(\theta)
   \left\{ \frac{3}{128\pi^2} \left[ 1 -2 \ln
   \frac{\mathring{M}^2(\theta)}{\mu^2} \right] + l_3^r + h_1^r - h_3 +
   l_7 \left[\frac{(1- \epsilon^2) \tan(\theta/2)}{ 1 + \epsilon^2\tan^2(\theta/2) } \right]^2
   \right\} ,
  \label{eq:evac}
\end{equation}
where we have dropped $\theta$-independent constant terms. The renormalized LEC
$l_3^r$ and HEC $h_1^r$ are scale dependent~\cite{Gasser:1983yg} and this scale
dependence cancels that in the chiral logarithm resulting in a scale-independent vacuum
energy density in a $\theta$-vacuum.
This is the main result of our paper.
It is then trivial to obtain the expression for any cumulant, and the lowest
two are
\begin{eqnarray}\label{eq:2.5}
\chi_t \al=\al \frac12 F^2 B \bar m \left(1-\epsilon^2\right)\left\{
  1 - \frac{2B \bar m}{F^2} \left( \frac{3}{32\pi^2}
  \ln\frac{2 B\bar m}{\mu^2} - 2 \left[l_3^r + h_1^r - h_3 - l_7
  \left(1-\epsilon^2\right) \right] \right) \right\} + \order{p^6},
  \nonumber\\
 c_4 \al=\al -\frac18 F^2 B \bar m \left( 1 + 2\epsilon^2 - 3\epsilon^4
 \right) + B^2 \bar m^2 \left(1-\epsilon^2\right) \left\{ \frac{9}{128\pi^2} \left(1-\epsilon^2\right)
   + \frac{3}{32\pi^2} \ln\frac{2 B\bar m}{\mu^2}\right. \nonumber\\
  \al\al - 2 \left[l_3^r + h_1^r - h_3 - l_7
  \left(1 + 2\epsilon^2 - 3\epsilon^4\right) \right] \bigg\}  + \order{p^6}.
\end{eqnarray}
They agree with the general $N$-flavor expressions in Ref.~\cite{Bernard:2012ci}
for $N=2$.
Furthermore, in the isospin symmetric case, they depend on the same combination
of the LECs and HECs, $l_3^r - l_7 + h_1^r - h_3$.

\section{SU($N$) with degenerate quark masses}

The evaluation of the functional determinant $Z_0(\theta)$ or
$e_\text{vac}^{(4,\text{loop})}(\theta)$ in Eq.~\eqref{eq:eloop} only requires
the Goldstone bosons to be degenerate.  Therefore, it is easy to generalize the
result in the previous section to the case of SU($N$) with degenerate quark
masses.\footnote{For SU($N$) with different quark masses, one may expand around
$\theta=0$, $\ln D_0(\theta) = \ln D_0(0) + D_0^{-1}(0)\,\Delta(\theta) +
D_0^{-1}(0)\,\Delta(\theta)\,D_0^{-1}(0)\,\Delta(\theta) + \ldots$ with
$\Delta_{PQ}(\theta)=
\delta_{PQ}\left[\mathring{M}_P^2(\theta)-\mathring{M}_P^2(0)\right]$. This
gives the general formulation used in Ref.~\cite{Bernard:2012ci}. } The one-loop
contribution to the vacuum energy density in a $\theta$-vacuum is given by Eq.~\eqref{eq:eloop} as well with $\mathring M(\theta)$ replaced by the LO Goldstone boson mass for SU($N$), see below.

When all the quarks are degenerate with a mass $m$, the vacuum is given by
$U_0=\mathbbm{1}_{N\times N}$.
With the $\order{p^4}$ Gasser--Leutwyler Lagrangian for
SU($N$)~\cite{Gasser:1984gg}, we get the tree-level contribution, including both
LO and NLO, to the vacuum energy density in a $\theta$-vacuum
\begin{equation}
  e_\text{vac}^\text{tree} = - N F_N^2 B_N m \cos\frac{\theta}{N} - 4 N B_N^2
  m^2 \left( 4 N L_6 \cos^2\frac{\theta}{N} - 4 N L_7 \sin^2\frac{\theta}{N} +
  2 L_8 \cos\frac{2\theta}{N} + 4 H_2 \right) ,
\end{equation}
where $L_{6,7,8}$ are LECs and $H_2$ is a HEC. Among them, $L_6, L_7$ and $H_2$
contain a UV divergent piece which can be calculated using the heat kernel
method with path integral~\cite{Gasser:1984gg,Bijnens:2009qm}
\begin{equation}
  L_6 = L_6^r + \frac{N^2+2}{16 N^2}\lambda\, , \qquad L_8 = L_8^r +
  \frac{N^2-4}{16 N}\lambda\, , \qquad H_2 = H_2^r + \frac{N^2-4}{8 N}\lambda
  \, .
\end{equation}
It is straightforward to check that these divergences cancel the one in
$e_\text{vac}^{(4,\text{loop})}$ in Eq.~\eqref{eq:eloop}. The vacuum energy
density in a $\theta$-vacuum up to NLO is then
\begin{equation}
  e_\text{vac}(\theta) = - \frac{N}{2} F_N^2 \mathring M_N^2(\theta) - \mathring
  M_N^4(\theta) \left\{ \frac{N^2-1}{128\pi^2} \left[1 -2 \ln \frac{\mathring
  M_N^2(\theta)}{\mu^2} \right] + 4 N\left( N L_6^r + L_8^r  - N
  L_7\tan^2\frac{\theta}{N} \right) \right\}
\end{equation}
with the scale-dependent finite LECs $L_6^r$ and $L_8^r$, where $\mathring
M_N^2(\theta) = 2 B_N m \cos(\theta/N)$, and the cumulants are
\begin{eqnarray}
  c_{2n} \al=\al \frac{(-1)^{n+1}}{N^{2n-1}} \left\{ F^2 B_N m + 4^n
  B_N^2 m^2 \left[ \frac{N^2-1}{64\pi^2 N}
  \left( 1 -2 \ln \frac{2 B_N m}{\mu^2} \right) + 8(N\, L_6^r + L_8^r + N\,
  L_7)  \right] \right\} \nonumber \\
 \al\al + \frac{N^2-1}{16\pi^2} B_N^2 m^2 \xi_{N,2n}
 \label{eq:c2n}
\end{eqnarray}
with the number $\xi_{N,2n}$ defined as
\begin{equation}
  \xi_{N,2n} = \left. \frac{d^{2n}}{d\,\theta^{2n}}\left[
  \cos^2\frac{\theta}{N}\, \ln\left(\cos\frac{\theta}{N}\right) \right]
  \right|_{\theta=0} .
\end{equation}
One sees that all cumulants depend on the same linear combination of the LECs,
as observed in Ref.~\cite{Bernard:2012ci} for the topological susceptibility
and the fourth cumulant, and chiral logarithms. From this it is easy to
construct LEC-free combination of cumulants which can be used for a clean
extraction of the $N$-flavor quark condensate from lattice simulations as
suggested in Ref.~\cite{Bernard:2012ci}. Examples are
\begin{eqnarray}
\chi_t + \frac{N^2}{4} c_4 \al=\al  \frac{3  F_N^2 B_N m}{4 N} +
\frac{3\left(N^2-1\right) B_N^2 m^2 }{32 \pi ^2 N^2} + \order{p^6} \, ,
\nonumber\\
\chi_t - \frac{N^4}{16} c_6 \al=\al  \frac{15  F_N^2 B_N m}{16 N} +
\frac{15\left(N^2-1\right) B_N^2 m^2 }{64 \pi ^2 N^2} + \order{p^6} \, ,
\label{eq:comb}
\end{eqnarray}
where the first expression was already proposed in
Ref.~\cite{Bernard:2012ci}\footnote{The physical pion mass was used in the
unitary logarithms in Ref.~\cite{Bernard:2012ci}. If one uses the LO pion mass,
one obtains agreement with the first expression here. The difference obtained
using the physical pion mass is of higher order. }.
More interestingly, the NLO corrections can be canceled out completely in
certain linear combinations, and lead to sum rules between
the QCD topological sector and the spontaneous breaking of chiral symmetry, such
as
\begin{equation}
  \Sigma_N = \frac{N}{m} \left( \frac{8}{5} \chi_t
  + \frac{2 N^2}{3} c_4 + \frac{N^4}{15} c_6 \right) + \order{p^6}.
\end{equation}
In fact, in the chiral limit, we have the following exact relation as can be
seen from Eq.~\eqref{eq:c2n}
\begin{equation}
  \Sigma_N = \pi\, \rho(0) = \lim_{m\to 0} (-1)^{n+1} N^{2n-1} \frac{c_{2n}}{m}
  \, ,
\end{equation}
where we have displayed the Banks--Casher relation~\cite{Banks:1979yr} linking
the quark condensate to the zero-mode spectral density of the Euclidean Dirac
operator, denoted by $\rho(0)$, as well. These relations can be simply obtained
using the LO expression for the vacuum energy density, and suggest that there is
an intimate link between the QCD topological sector and the spontaneous breaking
 of chiral symmetry.

\section{Summary}

We have derived the expressions for the vacuum energy density in a
$\theta$-vacuum in SU(2) CHPT up to NLO keeping different up and down quark
masses as well as in SU($N$) CHPT with degenerate quark masses. They can be used
to calculate the cumulants of the QCD topological charge distribution which are
important quantities to study QCD in the low-energy strong coupling regime.
In the case of degenerate quark masses, all cumulants depend on the same linear
combination of low-energy constants, as already observed for the topological
susceptibility and the fourth cumulant in Ref.~\cite{Bernard:2012ci}. Therefore,
one can construct many combinations of the cumulants depending only on the quark
mass and condensate. They can be used to extract the quark condensate in lattice
simulations without contamination from LECs. Furthermore, we find sum rules
relating the quark condensate to the cumulants free of NLO corrections. It would
be interesting to check such relations in lattice QCD.

\medskip

\section*{Acknowledgments}

We would like to thank V. Bernard, T.-W. Chiu, J. de Vries, M. D'Elia, S. 
Descotes-Genon,
K.
Ottnad and A. Rusetsky for useful discussions and comments. This work is supported in part
by DFG and NSFC through funds provided to the Sino-German CRC 110 ``Symmetries
and the Emergence of Structure in QCD'' (NSFC Grant No. 11261130311) and by NSFC
(Grant No. 11165005).

\end{document}